\RequirePackage[2020-02-02]{latexrelease}
\documentclass[notitlepage,twocolumn,prl,longbibliography,superscriptaddress]{revtex4-2}

\usepackage{amsmath}
\usepackage{changes}
\usepackage{graphicx}
\usepackage[space]{grffile} 
\usepackage{gensymb}
\usepackage{bm}
\usepackage{float}
\usepackage{amssymb,amsfonts,amsmath,amsbsy,bm,t1enc,latexsym}

\usepackage{changes}
\usepackage{soul}
\usepackage[super]{nth}
\usepackage[normalem]{ulem}
\usepackage[
bookmarks=true,
colorlinks=true,
linkcolor=blue,
urlcolor=blue,
citecolor=blue,
pdftex,
bookmarks=true,
linktocpage=true, 
hyperindex=true
]{hyperref}
\usepackage{hyperref}
\usepackage{siunitx}
\usepackage[super]{nth}
\usepackage[british]{babel}

\begin{document}

\title{Chaotic microcomb inertia-free parallel ranging}

	\author{Anton Lukashchuk}
	\affiliation{Laboratory of Photonics and Quantum Measurements (LPQM), Swiss Federal Institute of Technology Lausanne (EPFL), CH-1015 Lausanne, Switzerland}
	
	\author{Johann Riemensberger}
	\affiliation{Laboratory of Photonics and Quantum Measurements (LPQM), Swiss Federal Institute of Technology Lausanne (EPFL), CH-1015 Lausanne, Switzerland}	
	
	\author{Anton Stroganov}
	\affiliation{LIGENTEC SA, EPFL Innovation Park, CH-1024 Ecublens, Switzerland}
	
	\author{Gabriele Navickaite}
	\affiliation{LIGENTEC SA, EPFL Innovation Park, CH-1024 Ecublens, Switzerland}
	
	\author{Tobias J. Kippenberg}
	\email{tobias.kippenberg@epfl.ch}
	\affiliation{Laboratory of Photonics and Quantum Measurements (LPQM), Swiss Federal Institute of Technology Lausanne (EPFL), CH-1015 Lausanne, Switzerland}
	
	\date{\today}
	
	\pacs{}
	
	\maketitle

\textbf{
Ever growing pixel acquisition rates in the fields of augmented reality, autonomous driving and robotics have increased interest in solid state beam scanning without moving parts. Modern photonic integrated laser ranging advances towards passive beam steering solutions. Recently demonstrated imagers based on focal plane arrays, nanophotonic metasurfaces, and optical phased arrays enable unprecedented pixel resolution and measurement speed.
However, parallelization of >100 lasers and detectors -- successfully implemented in commercial time-of-flight sensors -- has not been widely adopted for passive scanning approaches.
Here, we show both inertia-free and parallel light detection and ranging (LiDAR) with microresonator frequency combs.
We used 40 independent channels of a continuously scanned microresonator frequency comb operated in the chaotic regime in combination with optical dispersive elements to perform random modulation LiDAR with 2D passive beam steering.
}

\section*{Introduction}

Laser beam steering is one of the key challenges in the LiDAR field. The ongoing search for passive scanning solutions ranges from focal plane grating arrays \cite{Zhang2022,Rogers2020} and switching networks \cite{Martin2018} to optical phased arrays \cite{Poulton2020} and nanophotonic metasurfaces \cite{Kim2021}.
Yet, conventional optical dispersive elements such as gratings and prisms offer mature and simple solutions for wavelength based beam steering. A combination of a virtually imaged phased array (VIPA) and a diffraction grating -- also referred as 2D disperser -- was demonstrated for 2D passive beam steering \cite{Bao2019,Goda2009} and frequency mapping into a 2D array in the image plane of the spectrometer for molecular fingerprinting \cite{Diddams2007}. The VIPA is a tilted etalon \cite{Shirasaki1996} that manifests angular dispersion by over an order of magnitude more than conventional diffraction gratings \cite{Xiao2004}. The subsequent mounting of a VIPA and a grating allows for vertical and horizontal passive scanning if the source frequency bandwidth is much larger than the free spectral range (FSR) of the VIPA etalon. 
Recently, Li et al. \cite{Li2021} demonstrated VIPA-based inertia-free frequency-modulated continuous-wave (FMCW) LiDAR with a broadband swept source.

Chip-scale optical microcombs can achieve 100s of nm of optical bandwidth and are suitable for massively parallel laser ranging either using FMCW \cite{Riemensberger2020,kuse_frequency-scanned_2021,shu2021sub} or random modulation continuous wave (RMCW) \cite{Lukashchuk2021} techniques. Such degree of parallelization enabled 6 megapixel per second line rates for FMCW microcomb LiDAR with one dimensional passive scanning \cite{Lukashchuk2022}. Yet, a demonstration of true passive 2D beam steering with optical microcombs was not explored.

Here, we report parallel and inertia-free microcomb based LiDAR.
We employ the recently demonstrated chaotic random modulation LiDAR approach \cite{Lukashchuk2021} where the photonic chip based microresonator is operated in chaotic modulation instability (MI) regime \cite{Matsko2013} to offer broadband noise, flat-top spectra, and high conversion efficiency. 
Based on the manifold of microcomb light and an optical disperser, we demonstrate passive and parallel 2D beam steering of a random modulation LiDAR system.

\begin{figure*}[!htbp]
\centering
	\includegraphics[width=0.92\linewidth]{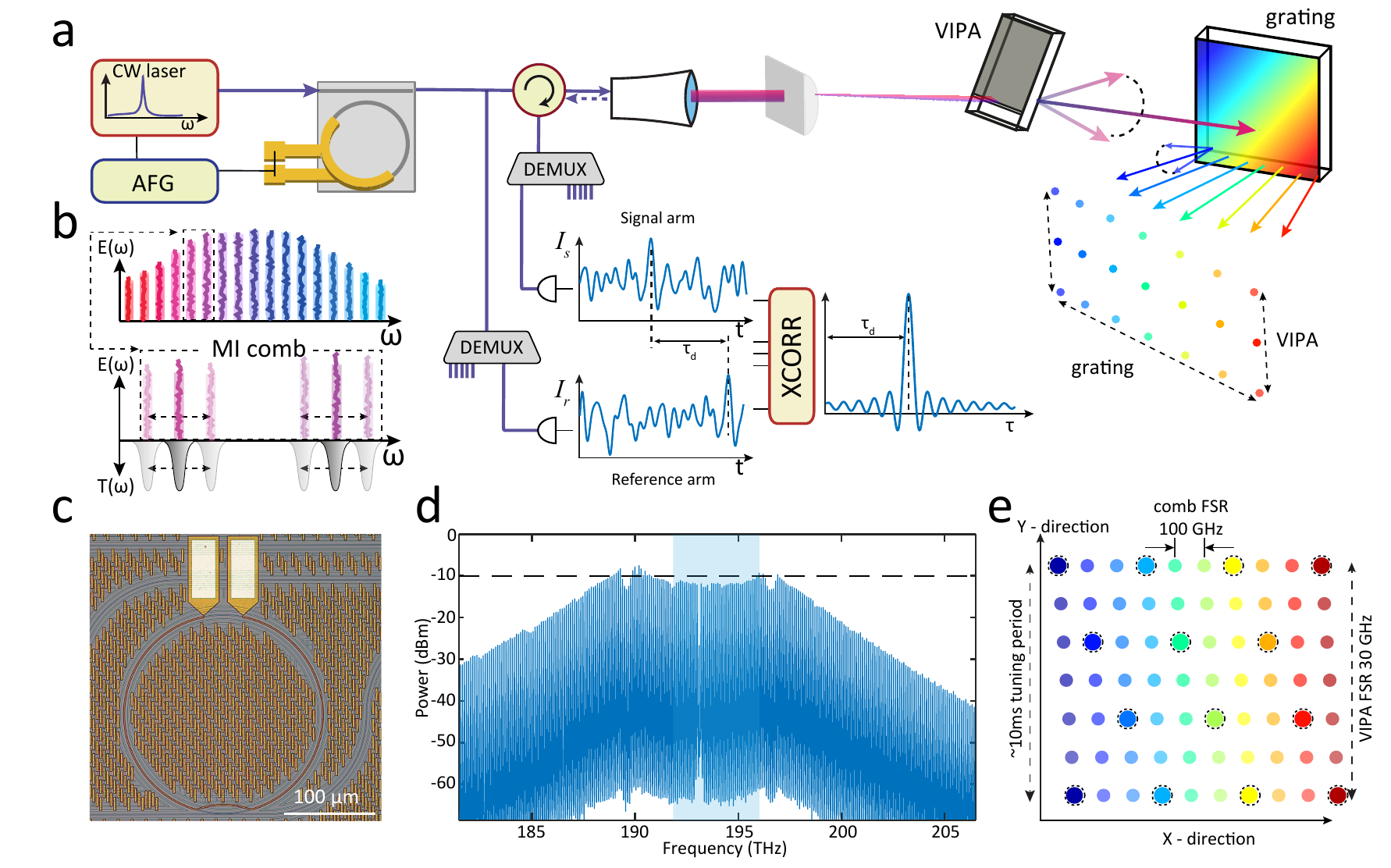}
	\caption{\textbf{Concept of chaotic microcomb based inertia-free ranging.}
	a)~Schematic of the experimental setup. The chaotic comb is generated with a continuous wave pump laser and 2D-spatially dispersed with a VIPA (virtually imaged phased array) and a diffraction grating. Each signal comb tooth represents an independent random LiDAR channel and is individually accessed via wavelength division demultiplexers. Cross-correlation of the reflected signal and the reference arm is used for range inference. An arbitrary function generator (AFG) simultaneously drives the piezo controller of the laser and the chip heater to scan the comb. 
	b)~Broadband scanning principle. The cavity resonances and comb lines are scanned in unison. Y-axis highlights the electric field amplitude E($\mathrm{\omega}$) and transmission T($\mathrm{\omega}$).
	c)~Microscope photo of the $\sim$227 $\mu$m radius Si$_{3}$N$_{4}$ microresonator with aluminum heater and contact pads.	
	d)~Optical spectrum of the chaotic comb with filtered out pump. The blue shaded region highlights comb channels (ITU 20-60) utilized for ranging experiments. The pump at 193 THz ($\mu$ = 0, ITU 30) is filtered.
	e)~2D spectral shower principle. The black dashed circles highlight a 2D mapping of the comb for a given fixed frequency offset. The diffraction grating disperses the comb in the X-direction while the VIPA of 30~GHz FSR disperses in Y-direction. The full vertical scan is achieved by tuning the comb through the VIPA FSR. If the comb FSR is not an integer of VIPA FSR, ccomb channels have different output angles proportional to \textit{mod($f_{\nu_0}+nFSR_{comb}$),FSR$_{VIPA}$)}.
	}
	\label{fig_concept}
\end{figure*}

\section*{Results and Methods}

\subsection*{Concept of chaotic inertia-free LiDAR}

The principle of inertia-free LiDAR with chaotic microcombs is illustrated in Fig.~\ref{fig_concept}a.
The photonic chip based integrated Si$_{3}$N$_{4}$ microring resonator \cite{Liu2021} driven by a continuous-wave (CW) laser generates a chaotic modulation instability frequency comb \cite{Godey2014} where each comb line acts as an independent random LiDAR source \cite{Lukashchuk2021, Lin2004}. Two photoreceivers per each optical channel detect the random-like varying light intensity and the cross-correlation between the reflected signal and the reference results in time delay estimate or range of a given pixel. A commercial wavelength demultiplexer unit enables individual comb line access and detection in parallel.

Two optical diffraction elements, VIPA and grating, realize 2D inertia free scanning. While the 100~GHz microcomb free spectral range (FSR) allows for beam diffraction by the grating in horizontal dimension, the VIPA with 30~GHz FSR is responsible for vertical scanning. Thus, the 2D disperser subsequently maps every frequency to a point in space. To leverage the angular dispersion of the VIPA, the microcomb needs to be scanned for 30~GHz. Simultaneous driving of the laser piezo controller and the microresonator heater achieves the required frequency excursion (similar to broadband soliton scanning \cite{Kuse2020}) keeping the frequency detuning $\Delta$ between the pump and the cavity resonance locked (cf.~Fig.~\ref{fig_concept} b). A description of the optical configuration of VIPA and gratin can be found in Ref.\cite{Xiao2004_2}.

We pump the chaotic microcomb with 900~mW of CW light at 1550~nm and achieve a 20$\%$ nonlinear conversion efficiency on-chip featuring up to $-$10 dBm comb line power and a flat-top spectrum (compared with a typical 'sech' profile of dissipative Kerr solitons \cite{Lukashchuk2021}). We have 100 comb lines with the mean power of -11.5~dBm and a standard deviation of 2~dB in the spectral region covering 188-198~THz. Each comb line has a characteristic GHz-level amplitude noise bandwidth, compared with 50~MHz resonance linewidth, resulting in $\sim$15~cm axial depth resolution and unambiguous ranging (due to the random-like waveform). The MI comb is generated blue-detuned from the optical resonance, i.e. it is thermally stable \cite{Brasch2016a}, and it follows a simple (compared with DKS) initiation routine. The blue shaded region on Fig. \ref{fig_concept}d highlights 40 comb channels (192-196~THz or ITU 20-60) utilized in the ranging experiments except for the filtered pump at 193~THz (ITU~30).  

The novel LiDAR scheme relies on simple direct detection with cross-correlation range inference, without the need for high frequency electro-optics or precise control over the optical waveform.  In our implementation, the microcomb scanning rate (shown up to 100~Hz) determines the overall frame rate of the proposed LiDAR since the field of view is covered twice during one scan.

\begin{figure*}[!htbp]
\centering
	\includegraphics[width=0.95\linewidth]{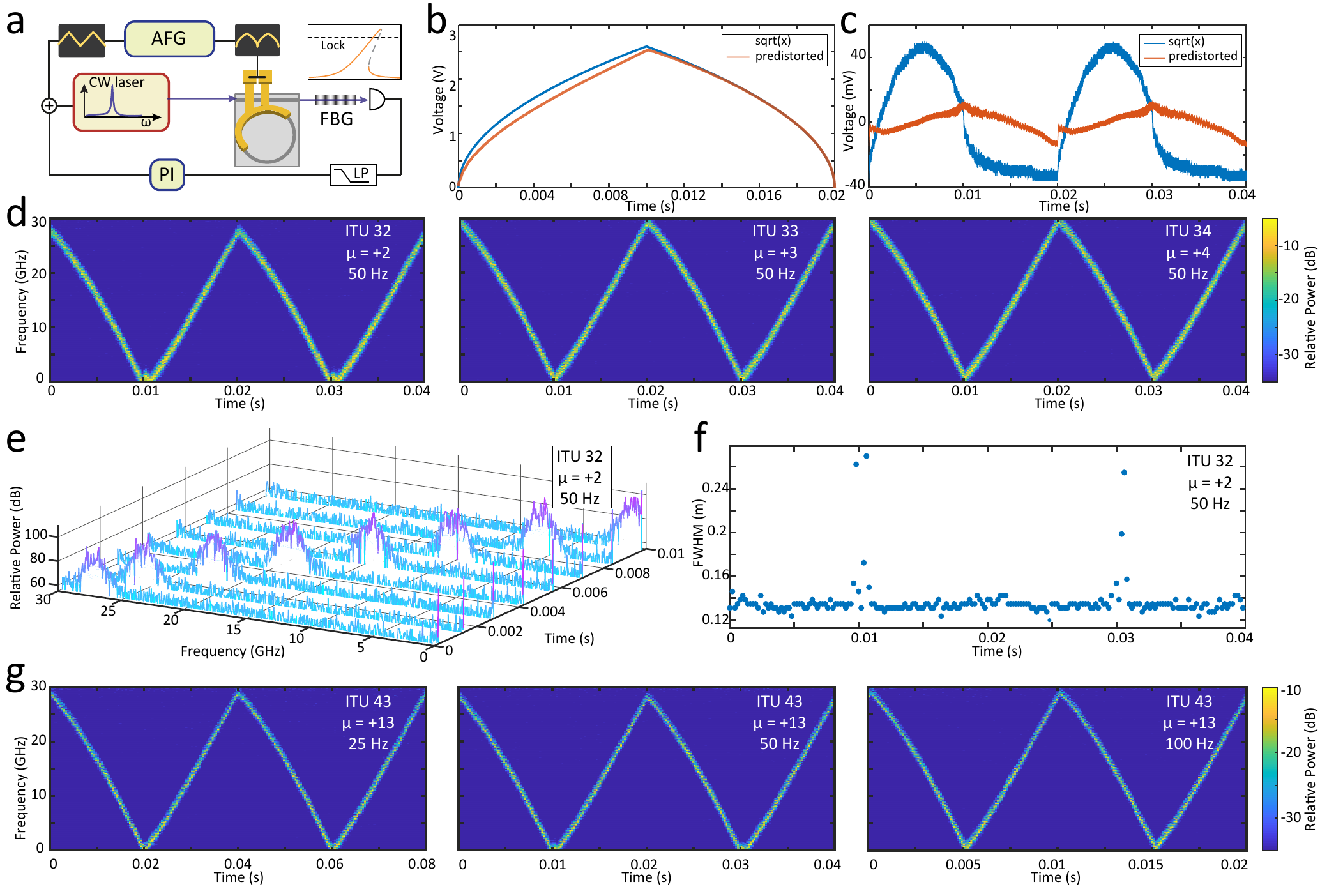}
	\caption{\textbf{Broadband chaotic microcomb scanning.}
	a)~Scanning principle. Square root and linear electrical waveform signals drive microresonator heater and laser piezo correspondingly. The laser is furthermore locked to the pump resonance via side of fringe locking (cf. top right) with PI feedback applied to the laser piezo in combination with the triangular waveform.
	b)~Initial square root-like (blue) and iteratively predistorted (red) waveforms applied to the heater. The error signal is derived from the photocurrent of the generated light (the pump is filtered).
	c)~The feedback error signal recorded on the oscilloscope during two periods for the heater waveforms described in b). Predistorted driving corresponds to triangular-like scanning and facilitates the subsequent lock.
	d)~Spectrograms of heterodyne detected $\mu$=2,3,4 comb channels at 50 Hz scanning. RBW = 14.6 MHz. 
	e)~Power spectrum evolution over half of the scan period ($\mu=2$). RBW = 29.1~MHz. Each spectra was recorded over 20 $\mu$s while continuous scanning as in d).
	f)~Full width half maximum of the auto-correlation traces ($\mu=2$) recorded over 20~$\mu$s per point while continuous scanning as in d). 
	g)~Heterodyne spectrograms for 25, 50, 100~Hz scan rates ($\mu$=13). RBW = 14.6~MHz
	}
	\label{fig_scan}
\end{figure*}

\subsection*{Sample fabrication}
We generate chaotic microcombs in Si$_3$N$_4$ optical microresonators fabricated using the photonic Damascene process \cite{Pfeiffer2018d}. A waveguide cross height of 800~nm ensures anomalous dispersion $D_2/2\pi = 700$~kHz at 1550~nm with intrinsic quality factor of 4 million. Deep-ultraviolet (DUV) stepper lithography was used to define structures and reactive ion etching and silica preform reflow \cite{Pfeiffer2018d} were done to reduce sidewall roughness. Aluminum-based integrated heaters were implemented using additional lithographic and metal etching steps, with the spacing from the waveguide defined such that their presence does not introduce any considerable optical loss. The optical tuning range of the microheater exceeds the 100~GHz FSR of the microring resonator, which allows for full spectral coverage of hte microcomb \cite{Kuse2020}.

\subsection*{Chaotic microcomb tuning}



Fig. \ref{fig_scan}a illustrates the MI microcomb scanning principle. We scanned both the pump laser and the optical microresonator to achieve a broadband frequency excursion covering the VIPA FSR of 30~GHz. The arbitrary function generator (AFG) drove the CW laser piezo controller with a triangular waveform and aluminum heater of the microresonator with a square root like waveform. The optical resonance shift is proportional to the temperature change (in the leading order) corresponding to the power dissipation or applied voltage squared. 
The broadband scanning of the microcomb also requires constant laser-resonance detuning $\Delta$ to maintain the MI state with uniform comb line power. We showed previously\cite{Lukashchuk2021} that the MI operated in pre-soliton switching regime carries the most intracavity power resulting in highest power per comb line and effective noise bandwidth resulting in better ranging resolution and accuracy.
To minimize the detuning $\Delta$ variation, we locked the pump to the slope of the resonance (side of fringe locking) via proportional-integral (PI) feedback control. We optimize the square root driving waveform of the integrated microheater by iterative predistortion to achieve linear (in the leading order) frequency chirping  (cf. Fig. \ref{fig_scan}b) and uniform MI comb power emission.

Fig. \ref{fig_scan}d depicts the heterodyne beat spectrograms of the $\mu$=2,3,4 comb lines  ($\mu$ stands for the comb line number relative to the pump) recorded by superimposing the comb lines with an external cavity diode laser. We detected the beatnote with a 43~GHz balanced photodiode and digitize the photocurrent with a high-speed oscilloscope (80~Gs/s) in segmented acquisition mode at a 50~Hz comb scanning rate and 10~$\mu$s segment length. The depicted spectrogram shows the Fourier transforms of the individual segments. 
The drift of the local oscillator during the measurement is less than 1 MHz. 
 The evolution of the MI comb line spectra for channel $\mu$ = 2 is depicted in Fig.~\ref{fig_scan}e. We observe a constant power and bandwidth of the chaotic beatnote. Fig.~\ref{fig_scan}f depicts the full width at half maximum (FWHM) of the auto-correlation traces derived from the scanned MI comb channel $\mu=2$. The 14~cm level determines resolution of the comb line at each step and confirms the locked detuning $\Delta$ (the reduced width around 0.01 and 0.03~s corresponds to the heterodyne spectrum folding around DC level). 

Fig. \ref{fig_scan}d depicts heterodyne spectrograms of the $\mu$=2,3,4 comb lines ($\mu$ stands for the comb line number relative to the pump) at 50~Hz rate. Second CW laser acted as a local oscillator (LO). High-speed 80~Gs/s oscilloscope recorded beatnotes on a broadband 43~GHz photodiode. The oscilloscope acquired the signal in segmented acquisition mode and 'smart' trigger setting. AFG trigger acted as an initial signal while the external clock acted as a subsequent trigger to acquire 100 segments of 10~$\mu$s each per period. Spectrograms compise Fourier transformed segments. The zoom-in in the spectrogram for channel $\mu$ = 2 shows MI spectra evolution  (Fig.~\ref{fig_scan}e).

In Fig.~\ref{fig_scan}g, we show the chaotic microcomb tuning at different tuning rates on the example of channel $\mu=13$. The fastest tuning rate that we could achieve is 100~Hz and it was limited by the response of the pump laser piezo controller. The aluminum heaters were reported to have response at kHz frequencies level \cite{Guo2021}, which could further increase the frame rate of the novel passive LiDAR.

\subsection*{Chaotic microcomb inertia-free ranging}
\begin{figure*}[!htbp]
\centering
	\includegraphics[width=0.92\linewidth]{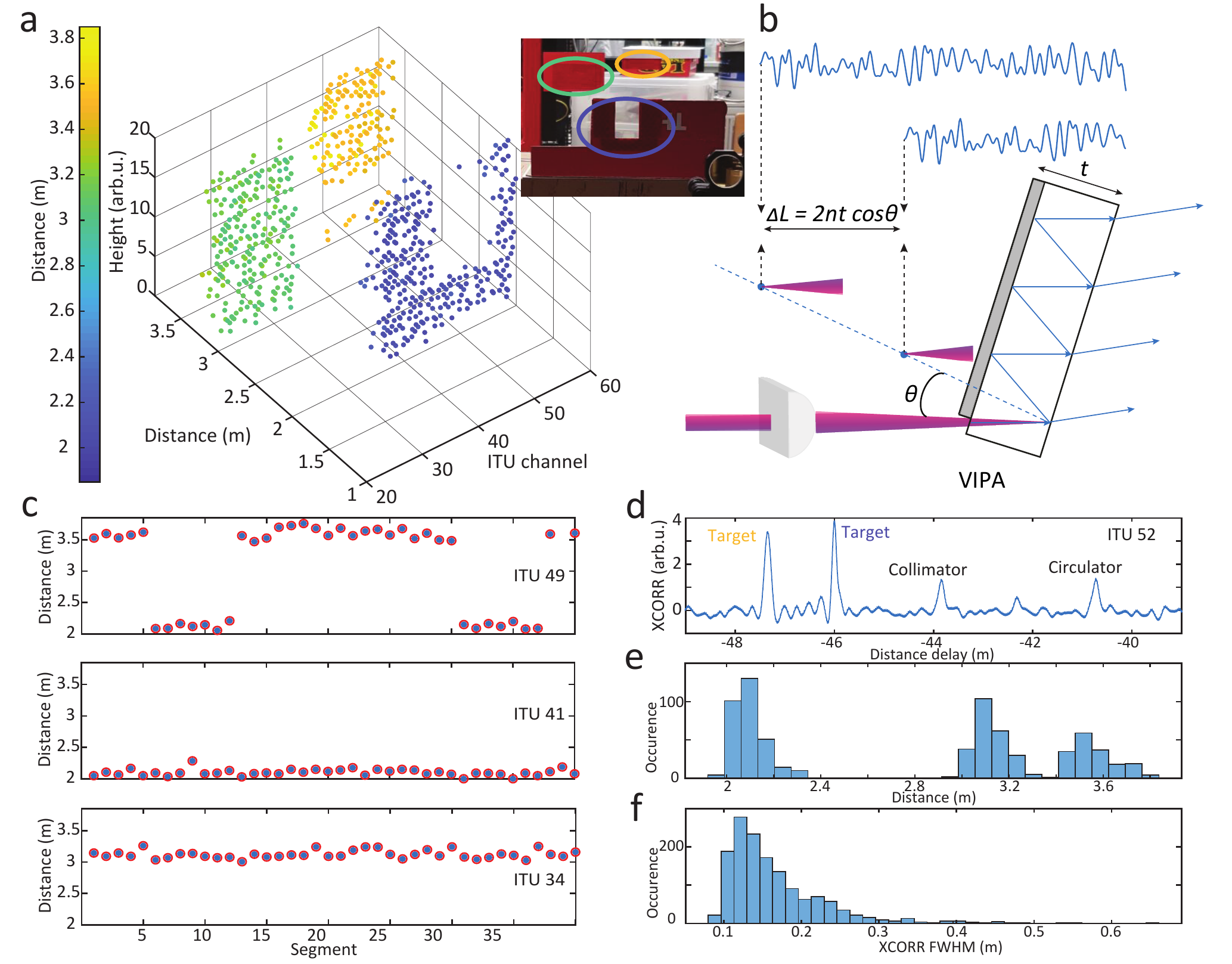}
	\caption{\textbf{Chaotic microcomb inertia-free ranging.}
	a)~Point cloud of 39$\times$20 pixels (top right: target photo). Height is depicted as 20 acquired segments corresponding to a full vertical VIPA swing over half of the scanning period.
	b)~Virtual imaged phased array principle. The drawback for RMCW or FMCW ranging with Fabry-Pero like disperer is shown as multiple sources illuminating the same waveform with different delays. The cross-correlation with a reference would result in a lower precision due to multiple delayed copies of the source. 
	c)~Ranging results for individual comb channels (ITU 49, 41, 34). 40 segments correspond to one scan period (0.02~s) with each segment recorded over 20~$\mu$s. 
	d)~Cross-correlation trace of one segment (20~$\mu$s acquisition) highlighting multiple targets. Two target reflections correspond to an edge case where VIPA operates in two orders.
	e)~Detection histogram highlighting $\sim$14~cm level precision.
	f)~Statistics of the target cross-correlation FWHM depicting $\sim$14~cm resolution.
	}
	\label{fig_ranging}
\end{figure*}

Fig. \ref{fig_ranging}a presents a point cloud of 39$\times$20 pixels acquired during 0.01~s - full vertical swing or half of the scanning period (50~Hz). 
We sequentially detected 39 comb channels (ITU 20-60 excluding filtered pump channel) while the detection can be performed in parallel requiring a number of photodetectors and digitizers equal to twice the number of channels (signal and reference for each channel). Every channel recording consisted of 40 segments (20~$\mu$s each) corresponding to two vertical scans (up-swing and down swing, cf. Fig.~\ref{fig_ranging}c; recorded in the segmented acquisition mode of the oscilloscope similar to heterodyne analysis Fig.~\ref{fig_scan}b). The demonstrated pixel measurement rate was limited by the available oscilloscope memory and the tuning rate of the external cavity diode laser. We used 1.5~GHz photoreceivers to record the chaotic intensity profiles and digitally computed cross-correlations for delay estimation. Cross-correlation trace comprises the target distance information and also the back-reflection from the collimator and circulator (cf. Fig.~\ref{fig_ranging}d). The dispersed comb illuminated carton blocks covered by retroreflecting tape at 2 to 3.6~m range away from the collimator. We achieve $\sim$14~cm depth resolution as analyzed from the cross-correlation FWHM (cf. Fig.~\ref{fig_ranging}f and Fig.~\ref{fig_scan}f). This is a direct consequence of a chaotic MI nature, where every comb line possesses >1~GHz effective noise bandwidth without the need for external modulation or control. Furthermore, the chaotic LiDAR is capable of unambiguous range detection \cite{Horton1959} and has superior immunity to ambient light interference \cite{Hwang2020}.

The 2D disperser scanning as depicted in Fig.~\ref{fig_concept}a corresponds to the case when the comb FSR equals to the VIPA FSR times an integer number. In our case - the VIPA and microcomb had 30~GHz and 100~GHz FSR, respectively - every third comb line illuminated with a similar vertical angle since the accumulated frequency separation of three FSRs was ten times the VIPA FSR (3$\times$100~GHz = 10$\times$30~GHz), while every subsequent comb line had a vertical angular shift - $\sim$33$\%$ of the vertical field of view (cf. Fig. \ref{fig_concept}e). Precise knowledge of the comb FSR and approximate VIPA FSR allowed us to calculate the relative vertical illumination angle shift for every comb line. The point cloud z-axis has arbitrary unit height value since we did not calibrate the absolute vertical angle corresponding to every segment of the linear sweep, though it is assumed to be linear in the leading order \cite{Xiao2004}. We used a high-density reflective diffraction grating  with 1200~lines/mm to diffract the light in the horizontal direction.

Fig. \ref{fig_ranging}e depicts a point cloud pixels histogram. The distance precision is on the order of the Fourier-limited range resolution ($\sim$14~cm) due to the multiple beam round-trips in VIPA etalon, i.e. multiple reflections with different accumulated delays (cf. Fig. \ref{fig_ranging}b) that can reach >10~cm with VIPA thickness of 3~mm and finesse $\mathcal{F}$ of 100. Another potential drawback may come from the effective frequency noise bandwidth being comparable to the VIPA linewidth, i.e. ratio of VIPA FSR and its finesse $\mathcal{F}$, which results in extra angular divergence of a particular chaotic comb line, even when the comb is not scanned.

The comb was amplified up to 500~mW prior to collimator.  Since we used fiber coupled photoreceivers, the returned light was projected into a fiber mode resulting in significant intensity loss due to speckle, which necessitated the use of retroreflector and pre-amplifier to overcome the digitizer noise level ($\sim$5~mV) at the detection stage.

\section*{Discussion}

We demonstrated chaotic microcomb based parallel inertia-free laser ranging. The proposed implementation requires minimum active components on the light transmitter side. Photonic integrated Si$_{3}$N$_{4}$ microresonator acts as a source of 40 and potentially >100 noisy comb channels for random LiDAR that are dispersed in 2D and can be detected in parallel. We showed that integrated aluminum micro-heaters enable agile frequency scanning of chaotic MI with 30~GHz excursion at 100~Hz rates (both limited by the laser piezo response) while Si$_{3}$N$_{4}$ microresonators were reported to achieve $\sim$200~GHz excursion and up to kHz scanning rates \cite{Kuse2020,Guo2021}. The higher excursion allows for VIPA with larger FSR and subsequently larger angular dispersion while higher scan rates translate into faster acquisition frame rates. The recently reported integration of aluminum nitride and lead zirconate titanate with Si$_{3}$N$_{4}$ photonic circuits \cite{Lihachev2022} could provide tuning with flat response and higher actuation bandwidth, but would require much higher voltage than aluminum heaters to cover several gigahertz range.
We note that heterogeneous integration of Si$_3$N$_4$ frequency comb with InP/Si semiconductor laser \cite{Xiang2021} has been demonstrated as well as hybrid integrated solutions \cite{Stern2018,Raja2019} and on-chip optical amplifiers \cite{Liu2022} paving the path to fully integrated LiDAR transmitter.

The combination of a VIPA and grating represents a simple yet mature 2D inertia-free scanning solution. It could provide 1.9$\times$7.7$^{\circ}$ field of view \cite{Li2021}, while the grating field of view could be easily extended by employing more comb lines or frequency combs with higher FSR. The VIPA tilt angle should be carefully chosen as it trades-off single order operation versus angular dispersion and tuning curvature \cite{Xiao2004}. 
Ultimately, any combination of passive optical dispersers can be considered. Non-etalon-based optical dispersers would eliminate the drawback of the multiple reflections and improve the ranging precision.

The sampling rate performance of the current scheme depends on the comb scanning period, pixel acquisition time and VIPA finesse. The total number of acquired pixels for one comb channel during one vertical scan should be less than the VIPA finesse $N<\mathcal{F}$. But also the total acquisition time should be less than the scanning period $ N \times \Delta t < T/2$. For 100~Hz tuning rate and a VIPA finesse $\mathcal{F} = 100$, the achievable sampling rate per comb channel is 20~kS/s at 50~$\mu$s pixel acquisition time. Megapixel sampling rates are feasible when more than 50 comb lines employed, or 50$\times$100 pixels at 200 frames per second rates.

Passive scanning is currently adopted as a commercial solution for true solid state LiDARs. Several LiDAR companies combine a 1D spectral scan approach based on optical disperses with mechanical scanning for the second dimension \cite{Okano2020, Baraja2020}. In addition, parallel illumination and acquisition - widely employed in time-of-flight LiDAR - may solve the long standing challenge of detecting megapixel rates required for real-time applications in robotics, unmanned driving and augmented reality. 
Finally, we have demonstrated parallel and inertia-free beam steering LiDAR based on chaotic microcombs.


\section{Funding}
This work was supported by the European Space Technology Centre with ESA Contract No 4000133568/20/NL/MH/hm, by the Air Force Office of Scientific Research under award number FA8655-21-1-7064. This work was further supported by the Swiss National Science Foundation (SNSF) under grant agreements No. 192293 and 201923 (AMBIZIONE).

\section{Acknowledgments}
AL thanks Arslan Raja for packaging efforts, Nikolai Kuznetsov for taking the microscope image Fig. \ref{fig_concept}c, Yang Liu for useful comments, Antonella Ragnelli and Kathleen Vionnet for administrative support. 

\section{Disclosures}
The authors declare no conflicts of interest.

\section{Data Availability Statement}
All data, figures and analysis code will be published on Zenodo upon publication of the work.


\bibliographystyle{apsrev4-1}
\bibliography{citations}

\end{document}